\documentclass[12pt,epsfig]{iopart}
\usepackage{epsfig}
\usepackage{epsfig}
\begin{document}

\title[Vegard's law for binary alloyed nano clusters]{ Mn$_m$Tc$_n$ nanoalloy clusters obey Vegard's law : A first principles prediction}
\author{Soumendu Datta and Tanusri Saha-Dasgupta}
\address{Department of Condensed Matter Physics and Material Sciences, S.N. Bose National Centre for Basic Sciences, JD Block, Sector-III, Salt Lake
City, Kolkata 700 098, India}
\date{\today}

\begin{abstract}
With a view to gain an understanding about the alloying tendency of bimetallic nano alloy clusters of isoelectronics constituents, we studied the structural and mixing behaviors of Mn$_m$Tc$_n$ alloy clusters with $m+n =$13 for all possible compositions, using first principles electronic structure calculations. Our study reports a favorable mixing tendency for the alloy clusters. The average bond lengths of the minimum energy structures show an overall linear variation with concentrations, indicating a Vegard's law like variation for the nano alloy clusters, though the optimized structures undergo a structural transition from a closed and compact structure for the Mn-rich alloy clusters to an open layered like structure for the Tc-rich alloy clusters.  We figure out a continuous and smooth interplay between hybridization and magnetization properties of the alloy clusters, which plays a vital role for the Vegard's law like variation in their average bond lengths. 
\end{abstract}


\section{\label{sec:intro}Introduction}

Bimetallic nano particles have shown immense applications, specially  in  catalysis as well as electronics, magnetic data storage devices and other cluster-assembled materials.\cite{alloy_rev1,alloy_rev2,alloy_rev3} During the last decade, nano sized bimetallic alloy clusters have been investigated extensively due to the progress in synthesis techniques to fabricate systems with well defined structures and controlled properties by varying size, concentration and chemical orderings of the constituent atomic species. For a nano alloyed system, the study of the surface structures, compositions and segregation properties are of particular interest  as they are the  main influencing factors  in determining the possibility of a surface chemical reaction. Therefore, a large fraction of these studies have been devoted to search the adopted structural patterns (core-shell versus randomly mixed) and the influencing physical parameters responsible for each particular pattern. The outcomes of these studies indicate that the degree of segregation/mixing and also atomic ordering of a bimetallic nanoalloy, vary from a system of one pair of species to another depending upon factors like, relative  atomic sizes, electronegativity difference of the constituent  species, surface energies of the bulk phases of the constituents, relative strengths of the associated homo-nuclear versus hetero-nuclear bonds\cite{alloy_rev1} etc. For example, small sized pure Cu$_n$, Ni$_n$ and Ag$_n$ clusters adopt disordered compact 3D-like structures,\cite{pure2,pure3} while the Cu-Ag and Ni-Ag nano alloys are reported to form core-shell like structures with larger atoms segregating on the surface and the smaller one at the core.\cite{cuag1,cuag2} Similarly, the previous studies reveal that pure Pd$_n$ clusters prefer a clear 3D structure and pure Au$_n$ clusters favor planar configurations,\cite{pure2,pure3} while the geometry of bimetallic  Pd$_m$Au$_n$ clusters can adopt either geometry, depending upon the relative concentrations of the constituents.\cite{pdau} Similar behavior is also reported for the other bimetallic nano systems like Fe$_m$Rh$_n$,\cite{ferh} Pt$_m$Fe$_n$\cite{ptfe} and so on. Another interesting fact to note about nano alloy clusters, is that several recent works report validation of Vegard's law like behavior for some bimetallic nano clusters, while deviation from it has also been reported. Examples includes the Au-Pt,\cite{aupt1,aupt2} Si-Ge,\cite{sige1,sige2} and Pd-Au\cite{pdau2} nano particles for which average bond lengths show a Vegard's law type behavior independent of their structural morphology, while the systems like Sn-Ge,\cite{snge} Fe-Mo/W\cite{fem} have lattice constants deviating from  Vegard's law behavior. The Vegard's law which states that  the lattice constant in a bulk binary alloy results from linear interpolation between the lattice constants of the pure constituent elements,\cite{vegard} is usually known as a bulk characteristic. Despite the fact that nano systems are expected to exhibit vastly different properties compared to its bulk counterpart, this observation of validation of Vegard's law like behavior for some nano alloy systems, is therefore an important finding. While Vegard's law for bulk binary systems, has been studied extensively in the past, its validation/departure for binary nano alloys is system dependent as mentioned above and not completely understood yet. Further study in this direction is, therefore, very desirable.

Interestingly, many of the above mentioned binary nano alloy systems for which the contrasting behaviors prevail, consist of elements belonging to a same column of the Periodic table and therefore, are isoelectronic. Several recent works on binary alloy clusters,\cite{iso1,iso2,iso3,iso4} studied the structures and relative stability of bimetallic alloy clusters consisting of two isoelectronic atom species and analyzed the effects of size mismatch and compositions. In the present study, we take up the case of MnTc alloy clusters in particular and study the aspect of their mixing tendency.

 The group VIIB transition metal element Manganese is an important magnetic element in molecular magnets, dilute magnetic semiconductors as well as a component in magnetic materials.\cite{dms1,dms2,dms3,magmat} Structure, magnetic and electronic properties of elemental small Mn clusters and bimetallic alloy clusters involving Mn atoms have been studied recently both experimentally as well as theoretically for searching novel nano magnetic materials. As for examples, the recent Stern-Gerlach deflection experiments reveal that while the small pure Mn$_n$ clusters with $n \ge 5$, have small net magnetic moments,\cite{mn1} the small Co$_m$Mn$_n$ clusters have been reported as superparamagnets with average per atom moments increasing with Mn concentrations and this enhancement is independent of cluster size.\cite{mnco1,mnco2} The small Bi-Mn clusters, on the other hand are found to induce either ferromagnetic or ferrimagnetic coupling behavior among the Mn atoms, depending on cluster size and composition.\cite{bimn} Moreover, some theoretical analysis on structure and segregation properties of Ti-Mn,\cite{timn} Fe-Mn,\cite{femn} Ni-Mn,\cite{nimn} Pt-Mn,\cite{mnpt1,mnpt2} and Au-Mn\cite{mnau1,mnau2} bimetallic alloy clusters have also been reported. The 4$d^5$ element Technetium which is the only radioactive transition metal element, is isoelectronic with Mn.\cite{tc} Recently, the possibilities of binary alloy formation for Tc have also been attempted to search for potential candidates for long-term nuclear-waste disposal in geological repositories.\cite{tc_alloy,tc_alloy2,tc_alloy3} Our study here, carried out on the structure and mixing tendency of Mn$_m$Tc$_n$ ($m+n$ = 13) alloyed clusters with varying compositions using first principles density functional theory (DFT) based electronic structure calculations, predicts a strong mixing tendency for the Mn$_m$Tc$_n$ alloy clusters with the cluster of maximum stability having nearly 50:50 composition. Further, the bond lengths of the optimized structures of the alloy clusters, show a Vegard's law like dependence on the Tc-atom concentrations. Detailed investigation of the electronic structure, provides a microscopic understanding for such variation. We believe that our analysis presented here, will help progressing the understanding of magnetic nanoalloy systems, in general.

\section{\label{methodology} Computational Details} The calculations reported in this study, were performed using DFT within the framework of pseudo potential plane wave method, as implemented in the Vienna abinitio Simulation Package (VASP).\cite{kresse2} We used the Projected Augmented Wave (PAW) pseudo potential \cite{blochl,kresse} coupled with the generalized gradient approximation (GGA) to the exchange correlation energy functional as formulated by Perdew, Burke and Ernzerhof (PBE).\cite{perdew} The 3$d$ as well as 4$s$ electrons for Mn atoms and 4$d$ as well as 5$s$ electrons for Tc atoms were treated as the valence electrons and the wave functions were expanded in the plane wave basis set with the kinetic energy cut-off of 280 eV. The convergence of the energies with respect to the cut-off value were checked. Reciprocal space integrations were carried out at the $\Gamma$ point. For the cluster calculation, a simple cubic super-cell was used with periodic boundary conditions, where two neighboring clusters were kept separated by around 12 {\AA} vacuum space, which essentially makes the interaction between cluster images negligible. Symmetry unrestricted geometry and spin optimizations were performed using the conjugate gradient and the quasi-Newtonian methods until all the force components were less than a threshold value of 0.001 eV/{\AA}. We considered both the collinear and non-collinear spin structures as detailed below, to 
achieve the minimum energy structure (MES) with the most relaxed spin configuration.

\begin{figure}
\rotatebox{0}{\includegraphics[height=10.1cm,keepaspectratio]{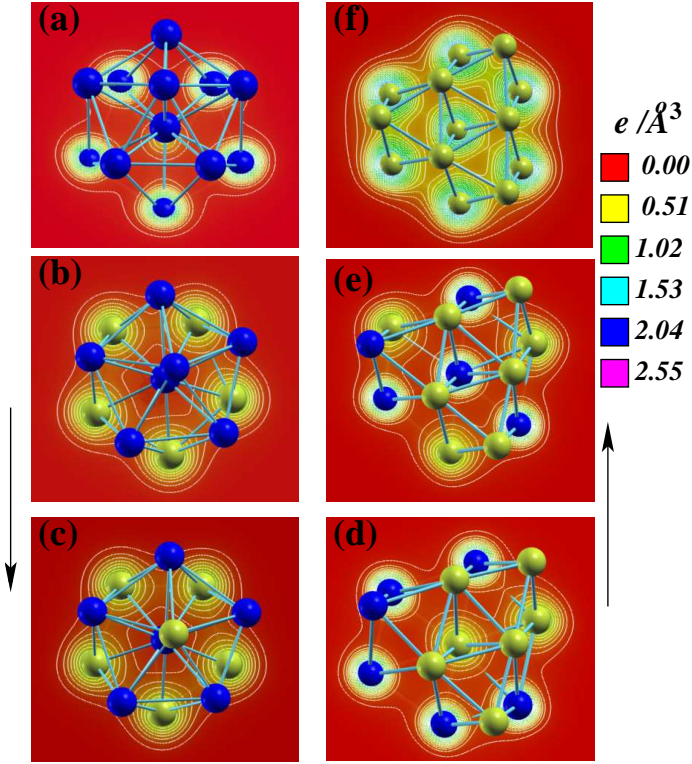}}
\caption{(Color online) MESs and the cross sections of the electronic charge density projected on to the xy plane for (a) Mn$_{13}$, (b) Mn$_8$Tc$_5$, (c) Mn$_7$Tc$_6$, (d) Mn$_6$Tc$_7$, (e) Mn$_5$Tc$_8$ and (f) Tc$_{13}$ clusters. The dark blue and light yellow balls represent 
Mn and Tc atoms respectively. The exhibited structures include the pure systems (top two panels) as well as the four compositions around which structural transition occurs (bottom four panels). The scale for charge density which is shown aside, is chosen to be the same across the 
varying compositions. The arrows point the systems of increasing Tc content.}
\label{mes}
\end{figure}

\section{\label{results}Results and Discussions}We first investigated the minimum energy structures, segregation as well as mixing tendency of the Mn$_m$Tc$_n$ alloy clusters. The favorable mixing tendency in the MESs of the Mn$_m$Tc$_n$ alloy clusters then prompted us to study their average bond length variation with Tc-concentrations. Interestingly, we observe an overall linear variation of the average bond lengths. Further analysis reveals interplay between the localized nature of the electronic states favoring magnetism and the delocalized nature of the electronic
states favoring hybridization energy gain, giving rise to a structural transition from compact to planar morphology, upon varying compositions 
which we identify as the responsible factor governing the linear variation of the bond lengths. The details of the results are given in various subsections. Section \ref{structure} summarizes the results about the MESs, the Section \ref{mixing} discusses the mixing and segregation tendency in the MES of the alloy clusters, and the Section \ref{microscop} analyzes the variation of average bond lengths in the MESs and provides the microscopic understanding for such variation.

\subsection{\label{structure}The MESs}
 We performed an exhaustive search for determining the MESs. Since the pure Mn$_{13}$ cluster shows a compact icosahedral (ICO) like structure and the pure Tc$_{13}$ cluster shows an open hexagonal bi-layered (HBL) structure as the MESs,\cite{iso4,avgr} we have taken these two geometries as initial guess morphologies to determine the MESs for the other compositions of the Mn$_m$Tc$_n$ clusters with $m$ + $n$ = 13. For the alloy clusters, first all possible distributions of the two species of atoms in these two morphologies ({\it i.e} homotops), have been considered to generate a large number of starting guessed structures. Then, we have optimized these guessed structures for each system, in non-collier calculations as well as collinear calculations with all possible spin multiplicities to ensure the lowest energy magnetic structure. We also considered different spin arrangements among the atoms of same concentration for a particular spin multiplicity. As a second step, we randomly displaced a few atoms in the optimized structures in order to take into account deviation from ICO like or HBL like geometry and to ensure the solution not getting stuck in the local minima in the complex potential energy surface, and re-optimized them. Fig. \ref{mes} shows MESs for the four selected alloy clusters, in addition with the MESs of the two pure extremes. It also shows the plot of their electronic charge density, projected onto the xy plane. It clearly indicates that while the Mn charges are mostly atom centered in the MES of the pure Mn$_{13}$ cluster, there is significant charge distribution in between the Mn and Tc atoms in the MESs of Mn$_m$Tc$_n$ alloy clusters, which is found to play an important role in their stability with implications in bonding behavior. Regarding the ordering of the two species of atoms, we found that the Tc atoms always prefer to sit at the surface sites of the icosahedron in case of Mn-rich alloy clusters, except for the case of Mn$_{12}$Tc cluster. For the MES of Mn$_{12}$Tc cluster, the only Tc atom likes to sit at central site which is understandable because of the small difference in their atomic sizes. For the MES of Mn$_8$Tc$_5$ cluster, the five Tc atoms prefer to occupy a pentagonal ring of the icosahedral structure. Likewise, the MES of the Mn$_7$Tc$_6$ cluster, continues to maintain the same structural motif of its predecessor, with the sixth Tc atom at the apex site farthest from the Tc-pentagon. In going from Mn$_7$Tc$_6$ to Mn$_6$Tc$_7$ clusters, a structural transition starts to occur from compact ICO-like structure to open HBL-like structure. 
The HBL-like motif becomes even more energetically favorable in the subsequent Tc-rich systems. The above conclusion is supported by calculated energy difference between the most optimized ICO and optimized HBL symmetry based structures for the all fourteen Mn$_m$Tc$_n$ clusters, which are : -1.45 eV, -0.95 eV, -1.16 eV, -0.69 eV, -0.45 eV, -0.27 eV, -0.12 eV, 0.04 eV, 0.29 eV, 0.82 eV, 1.49 eV, 1.72 eV, 1.93 eV and 3.70 eV starting from the Mn$_{13}$ cluster upto the Tc$_{13}$ cluster. The change in sign of the differences from -ve to +ve values, indicates a structural transition. It is also seen that the energy differences in general are more -ve as the Mn-concentration increases, while it is more +ve for the systems with increased Tc-concentrations. Regarding the total magnetic moments of the MESs, our calculations reveal an overall non-monotonic variation against the increasing concentrations of Tc atoms with the least value of around 1 $\mu_B$ for the MESs of Mn$_6$Tc$_7$ and Mn$_5$Tc$_8$ along with the MES of Tc$_{13}$ cluster. The three Mn-rich alloy clusters, namely Mn$_{12}$Tc, Mn$_{11}$Tc$_2$ and Mn$_{10}$Tc$_3$ though have relatively larger magnetic moments with the highest value of 19 $\mu_B$ in case of the MES of Mn$_{10}$ Tc$_3$ cluster. The other Mn$_m$Tc$_n$ clusters have some intermediate values such as 7 $\mu_B$ for the respective MES of Mn$_9$Tc$_4$, Mn$_8$Tc$_5$ and Mn$_3$Tc$_{10}$ clusters, 5 $\mu_B$ for the respective MES of Mn$_7$Tc$_6$ and Mn$_4$Tc$_9$ clusters and 3 $\mu_B$ for the respective MES of Mn$_{13}$, Mn$_2$Tc$_{11}$ and MnTc$_{12}$ clusters.

\subsection{\label{mixing}Mixing and segregation tendency}
One of the important quantities to study the segregation versus alloying behavior of a binary cluster, is the mixing energy,\cite{alloy_rev1,emix1} which is the difference in energy of the optimized alloy cluster from an appropriate fraction of energies of the identical configurations of its elemental constituents. For a 13 atom Mn$_m$Tc$_n$ alloy cluster, the mixing energy per atom is defined as
\begin{equation}
E_{mix}=\frac{1}{13}\left[E(Mn_mTc_n)-m\frac{E(Mn_{13})}{13}-n\frac{E(Tc_{13})}{13}\right]
\label{mee}
\end{equation}

\begin{figure}
\rotatebox{0}{\includegraphics[height=10.0cm,keepaspectratio]{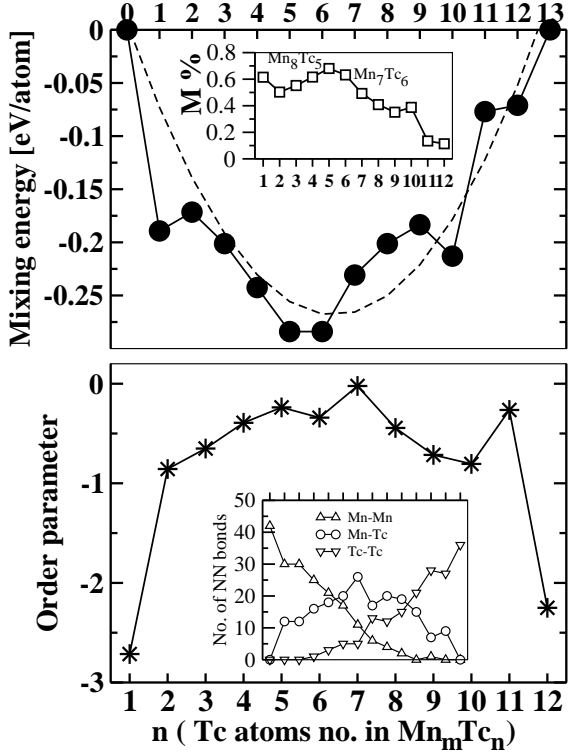}}
\caption{Variation of mixing energies (represented by solid dots in the top panel) and order parameters (represented by stars in the bottom panel) of the MESs of Mn$_m$Tc$_n$ clusters, as a function of number of Tc atoms in the clusters. The solid lines connecting the data points, are guide to the eye. The dashed line is the parabolic fitting through the mixing energy data. The inset in the top panel shows the variation of mixing coefficient of the MESs (by open squares) while the inset in the bottom panel shows variation of the total number of Mn-Mn NN bonds (by open up triangles), Mn-Tc NN bonds (by open circles) and Tc-Tc NN bonds (by open down triangles) as a function of Tc atoms. }
\label{me+op}
\end{figure}

Here, E(Mn$_m$Tc$_n$) is the energy for the MES of the Mn$_m$Tc$_n$ alloy cluster, while E(Mn$_{13}$) and E(Tc$_{13}$) are the energy of pure Mn$_{13}$ and Tc$_{13}$ clusters respectively in the same structure as of the MES for that composition. Eqn. (\ref{mee}) implies that the mixing energy is always zero for the pure extremes, while its negative values indicate that a heterogeneous phase formation in the alloy systems, will be favored i.e a tendency to form a nanoalloy out of Mn and Tc atoms and its smallest value will correspond to the most stable alloy cluster for this size. On the other hand, a positive value of mixing energy will characterize a segregation tendency among the constituent species. Fig. \ref{me+op} shows the plot of calculated mixing energies of all the MESs of Mn$_m$Tc$_n$ alloy clusters as a function of Tc atoms concentration in them. Interestingly, the mixing energies are negative for all compositions, indicating that the formation of heterogeneous phase in the Mn$_m$Tc$_n$ alloy systems for the whole range of compositions, is energetically favorable than separate phases of two species. The mixing energies first decrease (increasing stability) with increasing Tc-concentration in the Mn-rich alloy clusters, then it attains the minimum value for the MESs of two slight Mn-rich systems, namely Mn$_8$Tc$_5$ and Mn$_7$Tc$_6$ and finally it increases with the increase of Tc-atom concentrations in the Tc-rich alloy systems. This variation of mixing energies with composition, follows nearly a parabolic shape, as shown by the dashed line in Fig. \ref{me+op}. This analysis of mixing energy, shows that the Mn$_8$Tc$_5$ and Mn$_7$Tc$_6$ alloy clusters have gained particularly higher stability against segregation 
in comparison to the other compositions and therefore correspond to the magic compositions for the 13-atoms sized Mn$_m$Tc$_n$ clusters. It is also interesting to note from the Fig. \ref{mes} that the MESs of these two clusters have a close resemble in the sense that a unit of a sandwich like structure consisting of two pentagonal rings - one of 5 Mn atoms and another of 5 Tc atoms, connected by the central Mn atom, is common in both of them. The inset of the top panel of Fig. \ref{me+op} shows the variation of mixing coefficient M,\cite{alloy_rev1,BC} which is defined as percentage of the mixing energy in the total configuration energy i.e $M = \frac{E_{mix}}{E(m,n)}\times 100 \%$. It measures the relative contribution of mixing energy to the total energy. The highest peak for the Mn$_8$Tc$_5$ cluster and overall decreasing trend towards Mn-rich and Tc-rich systems, therefore, indicate that the mixing energy contribution to total energy is the maximum for the Mn$_8$Tc$_5$ cluster compared to the other compositions.

To understand the trend in mixing further, we studied the local chemical orderings of the constituent species in the MESs. To quantify it, we calculated a short ranged order parameter, as proposed by J. Cowley for bulk alloy systems\cite{op1} and verified recently for bimetallic nano alloys.\cite{op2} The order parameter for the Mn-Tc alloy clusters, is defined as
\begin{equation}
\alpha = 1 - \frac{N_{AB}/N}{\chi_{B}}
\label{formula_op}
\end{equation}
Here, $N_{AB}$ ($A,B$ = Mn, Tc) is the total number of nearest neighbor (NN) pairs between atoms of types $A$ and $B$. $N$ is the total number of nearest neighbor pairs which include all the Mn-Mn, Mn-Tc, Tc-Tc NN pairs. $\chi_B$ is the atomic concentration for atoms of minority type. Note that $\alpha \leq 0$ corresponds to mixed binary nano alloys, while $\alpha > 0$ for segregated nano alloy systems and $\alpha$ = 1 for pure systems. The bottom panel of the Fig. \ref{me+op} shows the variation of $\alpha$ versus number of Tc atoms in the alloy clusters. First note that all the alloy clusters have negative $\alpha$ values. Interestingly, the $\alpha$ values for all the alloy clusters except the singly doped systems, namely the Mn$_{12}$Tc and MnTc$_{12}$ clusters, fall within the range $-1 < \alpha < 0$, which is the range of homogeneous mixing as predicted earlier.\cite{op2} Therefore, this plot of order parameter reconfirms the trend towards mixing. For better understanding this trend in the variation of order parameter, the inset of the bottom panel shows separately the variation of the number of Mn-Mn, Mn-Tc and Tc-Tc NN bonds in the MESs with number of Tc-atoms. From the Eqn. \ref{formula_op}, it is seen that for similar morphologies (i.e Mn-rich or Tc-rich systems for which N is same), $N_{Mn-Tc}$ plays a significant role in the variation of $\alpha$ as  $N_{Mn-Tc}$ varies more rapidly than $\chi_B$. Therefore, as the number of Mn-Tc NN bonds increases with increasing Tc-content for the Mn-rich alloy systems, the order parameter becomes less negative. In going from Mn$_7$Tc$_6$ to Mn$_6$Tc$_7$ clusters, there is a change in $N$ values from 42 to 36 driven by the structural change. Thereafter, the overall effects of decreasing trend of N$_{Mn-Tc}$ and a decreased N value for the Tc-rich systems, restrict the $\alpha$ values also in the range $-1 < \alpha < 0$.

\begin{figure}
\rotatebox{0}{\includegraphics[height=8.6cm,keepaspectratio]{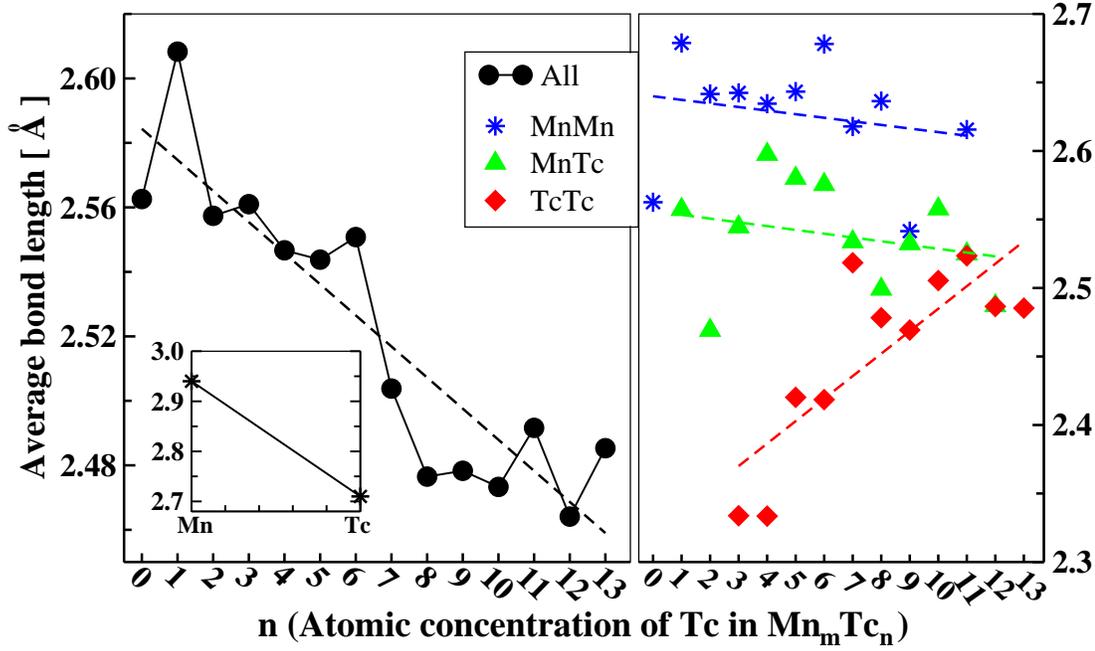}}
\caption{ (Color online) Variation of average bond lengths (represented by black colored solid dots in the left panel) and the average NN Mn-Mn, Mn-Tc, Tc-Tc (represented by blue colored stars, green colored solid triangles and red colored solid squares respectively) bond lengths with the concentrations of Tc atoms in the MESs of Mn$_m$Tc$_n$ clusters. The dashed lines are the linearly fitted curves for the corresponding sets of data. The inset of the left panel shows the linearly fitted curve between the Mn-Mn and Tc-Tc bond lengths of bulk $\alpha$-Mn and hcp Tc respectively.}
\label{abl}
\end{figure}

\subsection{\label{microscop}Variation of bond lengths and its origin}
Being motivated by the strong alloying tendency in the Mn$_m$Tc$_n$ alloy clusters, we moved forward to study the variation of average bond lengths with Tc concentrations for the MESs of the Mn$_m$Tc$_n$ alloy clusters. This is specially interesting to study as the Vegard's law for bulk binary alloy systems, is found to be valid mainly for randomly/continuously mixed alloy systems,\cite{veg1,veg2,veg3} as is the case in the present cluster systems. To calculate average bond lengths ($\langle d_{av}\rangle$) of the alloy clusters, we used the exponentially averaging weight function to take into account the contributions of the surrounding atoms at different distances, as followed in earlier works.\cite{avgr,iso4} Fig. \ref{abl} shows the plot of our calculated average bond lengths of the MESs for all compositions. As the NN bonds in case of a Mn$_m$Tc$_n$ alloy cluster, involve three types of bonds, namely Mn-Mn, Mn-Tc and Tc-Tc bonds, the right panel of Fig. \ref{abl} shows the respective variations of these NN bond lengths in the MESs, while the variation of overall average bond lengths with Tc concentrations is shown in the left panel. First it is important to mention that the inter atomic Mn-Mn bond lengths in bulk $\alpha$-Manganese is 2.25-2.94 {\AA},\cite{bulkmn} while Tc-Tc bond length in hexagonal close pack (hcp) lattice of bulk Tc, is 2.71 {\AA}.\cite{bulktc} Our calculated corresponding values for the MESs of pure Mn$_{13}$ and Tc$_{13}$ clusters, are 2.56 {\AA} and 2.48 {\AA} respectively. It is seen from the plot in the left panel of Fig. \ref{abl} that the average bond lengths decrease with the increasing Tc-concentrations. The interesting point to note is that this decrease of average bond lengths, follows roughly a linear interpolation between the average bond lengths of the two pure clusters as shown by the dashed line, in spite of the structural changes from ICO to HBL structures along the concentration variation as discussed in the Section \ref{structure}. The corresponding variation for the lattice constants of bulk MnTc alloys expecting according to Vegatd's law with respect to bulk Mn-Mn and Tc-Tc bond lengths in the respective pure bulk systems, would look like as shown in the inset of the Fig. \ref{abl}. Considering the variations of average bond lengths of the three types of NN bonds and the relative numbers of the respective NN bonds shown in the Fig. \ref{me+op}, one can expect the resulting trend in the overall average bond lengths variation. It is seen from the plot in the right panel of the Fig. \ref{abl} that the overall average NN Mn-Mn bond lengths  are larger than that of average NN Tc-Tc bond lengths with intermediate value of average NN Mn-Tc  bond lengths, in agreement with the expectation following bond lengths in the pure Mn$_{13}$ and Tc$_{13}$ clusters. The average NN Tc-Tc bond length though, shows a relatively faster variation in its value across the alloy series, being of smaller values for Mn-rich clusters and larger values from Tc-rich clusters, but still being smaller than the average Mn-Mn bond length. It is therefore obvious that both the numbers as well as the bond lengths of the three types of NN, play role in deciding the overall decreasing trend of the average bond lengths in the left panel of the Fig. \ref{abl}. While the number of Mn-Mn NN bonds dominate for Mn-rich clusters and that of Tc-Tc NN bonds dominate for Tc-rich clusters, the systems of intermediate concentrations have larger Mn-Tc NN bonds with the maximum value around 50:50 concentration.

To understand the trend in the average bond lengths variations, we look into the microscopic origins involving electronic properties. From electronic point of view, the valence $s$ and $d$ electrons of the constituents, are the main players. For a Mn atom, the energy level separation of the valence $s$ and $d$ levels is quite large,\cite{iso4} which indicates a small hybridization between the two, thereby keeping intact its high magnetic character because of half-filled $d$ valence shell. On the other hand, for Tc atom, this $s$ and $d$ levels separation is much smaller,\cite{iso4} which increases the $s$-$d$ hybridization, and pays by loosing magnetic character. 
We have quantified these two properties in terms of $s$-$d$ hybridization index and magnetic energy for the MES of each Mn$_m$Tc$_n$ cluster. 
\begin{figure}
\rotatebox{0}{\includegraphics[height=5.0cm,keepaspectratio]{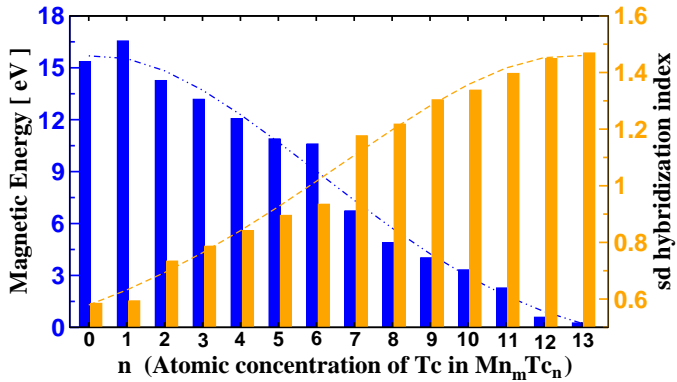}}
\caption{(Color online) Variation of magnetic energy (represented with respect to left y-axis) and sd hybridization index (represented with respect to right y-axis) with Tc-atom concentrations for the optimized structures of Mn$_m$Tc$_n$ clusters. The blue (dark) vertical bars correspond to the magnetic energies and the vertical orange (light) bars correspond to the sd hybridization index of the respective systems. The dashed lines are the fitted curves for the two sets of data.}
\label{interplay}
\end{figure}

The magnetic energy\cite{iso4,tseries} can be defined as the energy difference between the spin-polarized and non-spin polarized calculations for the optimal structure. Likewise, the $s$-$d$ hybridization index\cite{tseries} which measures the hybridization between $s$ and $d$ orbitals of the constituent atoms, can be defined as
\begin{equation}
 H_{sd}=\sum\limits_{I=1}^{13}\sum\limits_{i=1}^{occ}w_{i,s}^{(I)}w_{i,d}^{(I)}
\end{equation}
 where $w_{i,s}^{I}$ ($w_{i,d}^{I}$) is the projection of $i$-th Khon-Sham orbital onto the $s$ ($d$) spherical harmonic centered at atom $I$, integrated over a sphere of specified radius. The spin index is implicit in the summation. Fig. \ref{interplay} shows the variation of $s$-$d$ hybridization indexes and magnetic energies for the MESs of the Mn$_m$Tc$_n$ clusters for all the compositions. This plot displays an interesting interplay between the two quantities, with their smooth and continuous variations across the alloy series. The Mn-rich systems have higher magnetic energy gain and less hybridization index and vice-versa for the Tc-rich systems. The high magnetic energy gain for Mn-rich clusters, favors localization and therefore, result into larger inter atomic bond lengths. On the other hand, the enhanced hybridization energy gain for the Tc-rich alloy clusters, promotes delocalized character and manifests into lower inter atomic bond lengths.
This leads us to conclude that the smooth interplay between magnetization and hybridization and the continuous evolution of the electronic character
from localized to delocalized one, across the alloy series, is responsible not only for the structural transition between a compact and an open morphology,\cite{iso4} but also leads to a Vegard's law like behavior in the average bond length variation for the Mn$_m$Tc$_n$ alloy clusters with respect to the reduced Mn-Mn and Tc-Tc bond lengths in the respective pure systems.

\section{Summery and Conclusions}
Using first principles electronic structure calculations, we have studied the structures, stability, magnetism and mixing/segregation properties of Mn$_m$Tc$_n$ alloy clusters for all possible compositions with $m + n$ = 13. The alloy clusters show a favorable mixing tendency, as seen for other isoelectronic alkali binary clusters.\cite{iso1} The minimum mixing energy is located at the magic composition (m,n) $\equiv$ (8,5). Moreover, the average bond lengths for the optimized structures show a Vegard's law type variation with the increasing concentration of Tc-atoms, in spite of a structural transition from a compact icosahedral like structure for the Mn-rich clusters to an open HBL structure for the Tc-rich clusters. Our microscopic analysis reveals that a smooth and continuous interplay between $s$-$d$ hybridization and magnetization effects, is operative behind such variation of average bond lengths. Such analysis may also be carried out for other binary transition metal clusters, where bond lengths show a Vegard's law type variation.

\section*{Acknowledgments}

 S. D. thanks Department of Science and Technology, India for support through INSPIRE Faculty Fellowship, Grant No. IFA12-PH-27. T.S.D. would like to thank Department of Science and Technology for support.

\end{document}